\documentclass[twocolumn,aps,pra,amsmath,amssymb]{revtex4-1}
\usepackage[latin9]{inputenc}
\usepackage{amstext}
\usepackage{amssymb}
\usepackage{graphicx}
\usepackage{esint}

\makeatletter
\usepackage{amssymb}
\usepackage{mathbbold}
\usepackage{amsfonts}
\usepackage{epsfig}
\usepackage{bm}\usepackage{dcolumn}\usepackage{color}
\usepackage{overpic}
\usepackage[bottom]{footmisc}

\makeatother

\begin{document}

\title{Superfluid density reduction and spin-imbalanced pairing in a fermionic superfluid due to dynamical boson exchange}

\author{Ziyue Wang$^{1}$} 

\author{Lianyi He$^{1,2,3}$} 
\email{lianyi@mail.tsinghua.edu.cn}

\affiliation{$^{1}$ Department of Physics, Tsinghua University, Beijing 100084, China\\
$^{2}$ State Key Laboratory of Low-Dimensional Quantum Physics, Tsinghua University, Beijing 100084, China\\
$^{3}$ Collaborative Innovation Center of Quantum Matter, Beijing 100084, China}

\date{\today}
\begin{abstract}
We explore novel features of a nonrelativistic fermionic superfluid in which the pairing interaction includes a contribution from the exchange of a dynamical bosonic mode.  We show that the dynamical boson exchange (DBE), which causes a retarded pairing interaction and thus violates the Galilean invariance of the fermion sector,  generically leads to a quantum reduction of the superfluid density and hence a nonzero normal fraction even at zero temperature. For spin-singlet pairing, the DBE also leads to a nonvanishing spin susceptibility 
at zero temperature, providing a mechanism for the coexistence of pairing and magnetization.  While these effects are negligible for weak pairing, they become sizable at strong pairing. 
For the double superfluidity in ultracold Fermi-Bose mixtures,  the superfluid density reduction for the fermion sector induced by the DBE just gives rise to the Andreev-Bashkin drag effect, indicating a strong entrainment between the two superfluid components. The DBE may also provide a new source for the superfluid fraction reduction of neutron matter, which is crucial for models of neutron star glitches based on neutron superfluidity.
\end{abstract}

\pacs{05.30.Fk, 03.75.Ss, 67.85.Lm, 74.20.Fg}

\maketitle

\section{Introduction}

It is generally believed that an attractive interaction in a cold many-fermion system leads to superconductivity or superfluidity,  which covers a wide range of
many-fermion systems, including superconducting materials \cite{Review01}, superfluid $^3$He \cite{Review02}, ultracold atomic Fermi gases \cite{Review03}, excitonic condensate of electron-hole pairs \cite{Exciton}, nuclear matter \cite{Review04},  and dense quark matter \cite{Review05}. Ultracold atomic gases with tunable interatomic interaction provide clean systems to explore new phenomena 
associated with fermion superfluidity at strong coupling \cite{Greiner2003,Jochim2003,Zwierlein2003,Nascimbene2010,Horikoshi2010,Ku2012}. It has been demonstrated that 
a fermionic superfluid can evolve from a weakly paired BCS state to a Bose-Einstein condensate (BEC) of tightly bound pairs via increasing the attractive strength \cite{Eagles1969,Leggett1980,Nozieres1985,Melo1993,Chen2005,Gurarie2007}. 

According to Landau's two-fluid theory for a single-component superfluid \cite{Leggett-Book}, the superfluid density $\rho_s$ and the normal density $\rho_n$, satisfying $\rho_s+\rho_n=n$, with $n$ being the total particle density, characterize the superfluid and the normal components, respectively.  At the lowest temperature ($T=0$),  Galilean invariance leads to $\rho_s=n$ and hence 
$\rho_n=0$ \cite{Leggett-Book}. The violation of Galilean invariance naturally leads to a quantum reduction of $\rho_s$ \cite{Liang2017,Kinnunen2018,Watanabe2017,Huang1992,Orso2007,Zhou2012,He2012,Zhang2016,He2018}.  
At the one-body level, this may be realized by engineering the single-particle properties in cold atom
experiments, such as optical lattice \cite{Liang2017,Kinnunen2018,Watanabe2017}, disorder \cite{Huang1992,Orso2007}, and spin-orbit coupling \cite{Zhou2012,He2012,Zhang2016}. 
A quantum reduction of $\rho_s$ is yet to be explored in cold atom experiments.

In this work, we propose a more fundamental mechanism for a quantum reduction of the superfluid density at the two-body level.  We study a nonrelativistic fermionic superfluid in which the pairing interaction includes a part mediated by a dynamical bosonic mode.  Due to the dynamical boson exchange (DBE), the effective two-body interaction between fermions is not static. The interaction retardation effect thus violates the Galilean invariance for the fermion sector and leads to a quantum reduction of the superfluid density. We find that the reduction is vanishingly small in the weak pairing limit. However, for strong pairing, the quantum reduction becomes sizable. The boson-mediated pairing interaction can be realized in dilute Fermi-Bose mixtures, where the force carrier becomes the Bogoliubov phonon mode of BEC \cite{Heiselberg2000,Viverit2000,Stoof2000}. In the previous studies of fermion pairing in Fermi-Bose mixtures, the interaction retardation due to DBE was normally neglected \cite{Matera2003}. On the other hand, this two-component superfluid mixture should be described by the three-fluid theory \cite{Three-Fluid,Andreev1975}. We show that in this double-superfluid system, the quantum reduction of the fermionic superfluid density also gives rise to the Andreev-Bashkin drag effect \cite{Andreev1975}, as required by the Galilean invariance of the whole Fermi-Bose system.  A strong pairing thus leads to a strong entrainment between the two superfluid components.

The spin susceptibility $\chi$ is another important quantity for fermionic superfluidity.  For static pairing interaction, it vanishes at $T=0$ for spin-singlet $s$-wave pairing, indicating that the 
$s$-wave pairing is not compatible with a nonzero magnetization at $T=0$ \cite{Review06}.  In this case, the application of a Zeeman field does not induce a magnetization in the superfluid state, and a first-order phase transition to the normal state occurs at the so-called Chandrasekhar-Clogston or Pauli limit \cite{CC1962,FFLO1964, spin-imbalance}. One route to realize a nonzero $\chi$ at $T=0$ is to turn on the spin-orbit coupling \cite{Gorkov2001}, which induces a spin-triplet pairing component even though the pairing interaction is of the $s$-wave nature.
In this work, we show that for spin-singlet $s$-wave pairing,  the DBE also leads to a nonzero spin susceptibility even at $T=0$, providing a mechanism 
for the coexistence of pairing and magnetization.  The application of a Zeeman field induces a finite magnetization in the superfluid state.

\section{Fermionic superfluidity with boson exchange}

We consider a gas of spin-$1/2$ nonrelativistic fermions with bare mass $m_{\rm f}$ in free space. The interaction between fermions includes a part mediated by
a bosonic mode.  The Lagrangian density of the system can be given by ${\cal L}={\cal L}_{\rm f}+{\cal L}_{\rm b}+{\cal L}_{\rm bf}$. The fermion part reads
\begin{eqnarray}
{\cal L}_{\rm f}=\sum_{\sigma=\uparrow,\downarrow}\psi^\dagger_\sigma\left(-\partial_\tau+\frac{\nabla^2}{2m_{\rm f}}\right)\psi^{\phantom{\dag}}_\sigma+{\cal L}_{\rm ff},
\end{eqnarray}
where $\tau$ is the imaginary time, $\psi_\sigma$ represents the Grassmann field for fermion with spin $\sigma$, and ${\cal L}_{\rm ff}$ denotes the static two-body interaction between the fermions. We use the units $\hbar=k_{\rm B}=1$ throughout.  The boson part ${\cal L}_{\rm b}$ 
and the fermion-boson interaction ${\cal L}_{\rm bf}$ are not specified here.

Formally we may integrate out the bosonic mode and obtain an action with only fermions. The action reads
${\cal S}={\cal S}_0+{\cal S}_{\rm int}$,  with the single-particle part given by
${\cal S}_0=\sum_{\sigma}\sum_{K}(-i\omega_n+\xi_{\bf k})\psi^\dagger_\sigma(K)\psi^{\phantom{\dag}}_\sigma(K)$.
Here $K=(\omega_n,{\bf k})$ denotes the fermion Matsubara frequency $\omega_n=(2n+1)\pi T$ and 
momentum ${\bf k}$, and $\xi_{\bf k}={\bf k}^2/(2m_{\rm f})-\mu_{\rm f}$ is the free fermion dispersion, with $\mu_{\rm f}$ being the chemical potential.  
The interaction part can be expressed as
\begin{eqnarray}
{\cal S}_{\rm int}=\frac{1}{2\beta{\cal V}}\sum_{\sigma=\uparrow,\downarrow}\sum_{Q}V_{\sigma\sigma^\prime}(Q)\rho_\sigma(Q)\rho_{\sigma^\prime}(-Q),
\end{eqnarray}
where $\rho_\sigma(Q)=\sum_{K}\psi^\dagger_\sigma(K+Q)\psi^{\phantom{\dag}}_\sigma(K)$, $\beta=1/T$, ${\cal V}$ is the volume of the system, and $Q=(q_l,{\bf q})$ denotes the boson Matsubara frequency $q_l=2l\pi T$ and momentum ${\bf q}$. 
The frequency and momentum dependent effective interaction $V_{\sigma\sigma^\prime}(Q)$ includes the contribution from the exchange of the bosonic mode. Because of the
contribution from DBE, the effective interaction between the fermions is not static. The interaction retardation thus violates the Galilean invariance of the fermion sector.

We consider equal spin populations ($n_{\uparrow}=n_{\downarrow}$) and assume that the spin-singlet pairing dominates. We thus rewrite the relevant pairing interaction as
\begin{eqnarray}
{\cal S}_{\rm int}=\frac{1}{\beta{\cal V}}\sum_Q\sum_{K,K^\prime}{\cal B}_K^\dagger(Q)U(K-K^\prime){\cal B}_{K^\prime}^{\phantom{\dag}}(Q),
\end{eqnarray}
where ${\cal B}_K(Q)=\psi_\downarrow(Q-K)\psi_\uparrow(K)$ and $U(Q)\equiv V_{\uparrow\downarrow}(Q)$.
Following the standard field theoretical treatment, we introduce a pairing field $\Phi_{K}(Q)$ via the Stratonovich-Hubbard transformation, which satisfies the equation of motion 
$\Phi_K(Q)=(\beta{\cal V})^{-1}\sum_{K^\prime}U(K-K^\prime){\cal B}_{K^\prime}(Q)$. Integrating out the fermions, we obtain an effective action
\begin{eqnarray}
{\cal S}_{\rm eff}&=&-\beta{\cal V}\sum_Q\sum_{K,K^\prime}\Phi_K^*(Q)U^{-1}(K-K^\prime)\Phi_{K^\prime}^{\phantom{\dag}}(Q)\nonumber\\
&&-{\rm Tr}\ln \left\{{\bf G}^{-1}_{K,K^\prime}[\Phi,\Phi^*]\right\}.
\end{eqnarray}
Here the inverse of $U$ is defined as $\sum_PU^{-1}(K-P)U(P-K^\prime)=\delta_{KK^\prime}$. The inverse of the fermion Green's function ${\bf G}$ in the Nambu-Gor'kov representation
$\Psi(K)=[\psi^{\phantom{\dag}}_\uparrow(K), \psi^\dagger_\downarrow(-K)]^{\rm T}$ is given by
\begin{eqnarray}
{\bf G}^{-1}_{K,K^\prime}=\left[\begin{array}{cc}(i\omega_n-\xi_{\bf k})\delta_{KK^\prime} & \Phi_K(K-K^\prime)\\  \Phi_{K^\prime}^*(K^\prime-K) & (i\omega_n+\xi_{\bf k})\delta_{KK^\prime}\\ \end{array}\right].
\end{eqnarray}

At low temperature the pairing field acquires a nonzero expectation value. We consider a static and homogeneous superfluid state and write
$\Phi_{K}(Q)=\Delta(K)\delta_{Q,0}+\tilde{\Phi}_K(Q)$, where $\Delta(K)$ serves as the order parameter of superfluidity. The path integral over the fluctuation $\tilde{\Phi}$ cannot be accurately evaluated. 
Here we mainly consider $T=0$ and hence employ the mean-field approximation. It is believed that the mean-field theory describes correctly the BCS-BEC crossover at $T=0$ \cite{Eagles1969,Leggett1980,Nozieres1985,Melo1993,Chen2005,Gurarie2007}. 

In the mean-field theory,  the grand potential reads
 \begin{eqnarray}
\Omega[\Delta(K)]&=&-\sum_{K,K^\prime}\Delta^*(K)U^{-1}(K-K^\prime)\Delta(K^\prime)\nonumber\\
&&-\frac{1}{\beta {\cal V}}\sum_K\ln\det \left[G^{-1}(K)\right],
\end{eqnarray}
with the fermion Green's function given by
\begin{eqnarray}
G^{-1}(K)=
\left[\begin{array}{cc}i\omega_n-\xi_{\bf k} & \Delta(K)\\  \Delta^*(K) & i\omega_n+\xi_{\bf k}\\ \end{array}\right].
\end{eqnarray}
Minimizing $\Omega[\Delta(K)]$, we obtain the gap equation 
\begin{eqnarray}\label{GEQ}
\Delta(P)=-\frac{1}{\beta {\cal V}}\sum_{K}U(P-K)\frac{\Delta(K)}{\omega_n^2+{\cal W}^2(K)}.
\end{eqnarray}
The total fermion density $n_{\rm f}=n_{\uparrow}+n_{\downarrow}$ is given by
\begin{eqnarray}\label{NEQ}
n_{\rm f}=-\frac{2}{\beta {\cal V}}\sum_{K}\frac{(i\omega_n+\xi_{\bf k})e^{i\omega_n0^+}}{\omega_n^2+{\cal W}^2(K)}.
\end{eqnarray}
Here ${\cal W}^2(K)=\xi_{\bf k}^2+|\Delta(K)|^2$. 

For static pairing interaction, $U(P-K)$ reduces to $U({\bf p}-{\bf k})$ and thus the pairing gap depends only on the momentum. The Matsubara sum can be worked out analytically and the gap and number equations reduce to the BCS-Leggett mean-field description of the BCS-BEC crossover \cite{Leggett1980}. In the presence of DBE, Eqs. (\ref{GEQ}) and (\ref{NEQ}) should be solved simultaneously to determine the full frequency and momentum dependence of the gap function $\Delta(\omega_n,{\bf k})$ as well as the chemical potential $\mu_{\rm f}$, and thus constitute a type of Eliashberg theory \cite{Eliashberg1960,Abrikosov-book,Wang2006,Bulgac2009}. Here we emphasize that unlike the previous studies, the full momentum dependence should be maintained in order to approach strong pairing, as was first considered in \cite{Bulgac2009}.
For convenience, we define the Fermi momentum $k_{\rm F}$ and the Fermi energy $\varepsilon_{\rm F}$ via a noninteracting Fermi gas, i.e., $k_{\rm F}=(3\pi^2n_{\rm f})^{1/3}$
and $\varepsilon_{\rm F}=k_{\rm F}^2/(2m_{\rm f})$. The Fermi velocity can be defined as $v_{\rm F}=k_{\rm F}/m_{\rm f}$.

\section{Linear Responses}

The superfluid density and the spin susceptibility can be evaluated via the standard linear response method. The superfluid density $\rho_s$ 
characterizes the response of the system to an infinitesimal uniform superfluid flow ${\bf v}_s$, which amounts to being equivalent to a constant U$(1)$ vector potential ${\bf A}=m_{\rm f}{\bf v}_s$ 
\cite{He2006, Taylor2006, Brendan2017}. For an isotropic superfluid state, the change in the free energy reads $F({\bf v}_s)-F({\bf 0})=\frac{1}{2}m_{\rm f}\rho_s {\bf v}_s^2+O({\bf v}_s^4)$. 
For convenience we can use the grand potential $\Omega$ instead of the free energy $F=\Omega+\mu_{\rm f} n_{\rm f}$, since the difference can be shown to be beyond the order $O({\bf v}_s^2)$ 
\cite{He2006, Taylor2006}.

The grand potential at finite ${\bf v}_s$ can be obtained by replacing the fermion Green's function $G^{-1}(K)$ with
\begin{eqnarray}\label{GA}
G_{\bf A}^{-1}(K)=
\left[\begin{array}{cc}i\omega_n-\xi_{{\bf k}-{\bf A}} & \Delta(K)\\  \Delta^*(K) & i\omega_n+\xi_{{\bf k}+{\bf A}}\\ \end{array}\right].
\end{eqnarray}
Performing the Taylor expansion in ${\bf v}_s$, we obtain $\Omega({\bf v}_s)-\Omega({\bf 0})=\frac{1}{2}m_{\rm f}\rho_s {\bf v}_s^2+O({\bf v}_s^4)$.
The superfluid density $\rho_s$ can be expressed as $\rho_s=n_{\rm f}-\rho_n$, where the reduction or the normal density is given by
\begin{eqnarray}\label{RHOn}
\rho_n=\frac{2}{\cal V}\sum_{\bf k}\frac{{\bf k}^2}{3m_{\rm f}}Y({\bf k}).
\end{eqnarray}
The function $Y({\bf k})$ is defined as a Matsubara sum
\begin{eqnarray}\label{Yfunction}
Y({\bf k})=\frac{1}{\beta}\sum_n\frac{\omega_n^2-\xi_{\bf k}^2-|\Delta(\omega_n,{\bf k})|^2}{\left[\omega_n^2+\xi_{\bf k}^2+|\Delta(\omega_n,{\bf k})|^2\right]^2},
\end{eqnarray}
which cannot be analytically evaluated in the general case. For the normal state with $\Delta=0$, we can show that $\rho_n=n_{\rm f}$ and thus $\rho_s=0$. In the absence of DBE,  
the gap function does not depend on the frequency, i.e., $\Delta(\omega_n,{\bf k})=\Delta_{\bf k}$. In this case, the Matsubara sum can be evaluated to give the 
standard Landau formula $Y({\bf k})=(4T)^{-1}{\rm sech}^2\left(E_{\bf k}/2T\right)$,
where $E_{\bf k}=\sqrt{\xi_{\bf k}^2+|\Delta_{\bf k}|^2}$ is the standard BCS excitation spectrum. It is obvious that $\rho_n=0$ at $T=0$.

\begin{figure*}
\centering{}\includegraphics[width=1.0\textwidth]{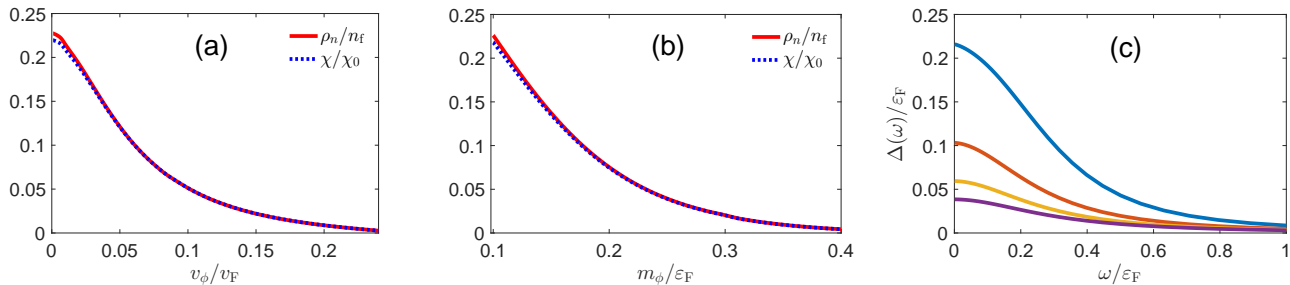} \caption{Results for the toy Yukawa model. (a) The dependence of $\rho_n$ (scaled by $n_{\rm f}$) and $\chi$ (scaled by $\chi_0$) 
on $v_\phi/v_{\rm F}$ at $m_\phi=0.1\varepsilon_{\rm F}$. 
(b) The dependence of $\rho_n$ and $\chi$ on $m_\phi/\varepsilon_{\rm F}$ at $v_\phi/v_{\rm F}=0$.
(c) The frequency dependence of the gap function at $v_\phi/v_{\rm F}=0$, for $m_\phi/\varepsilon_{\rm F}=0.1,0.15,0.2,0.25$ (from top to bottom). The Yukawa coupling is fixed at
$g^2/v_{\rm F}^3=0.1$.
\label{fig1}}
\end{figure*}


In the presence of DBE, the gap function depends on the frequency. In this case, $\rho_n$ does not vanish at $T=0$, which is a direct consequence of the violation of Galilean invariance induced by DBE.  To show this, we note that the Matsubara sum at $T=0$ is converted to an integral over
an imaginary frequency, $T\sum_n f(\omega_n)\rightarrow\int_{-\infty}^\infty d\omega/(2\pi) f(\omega)$. Using the trick of integration by parts \cite{Alford2018} shown in Appendix A, we can show that at $T=0$,
\begin{eqnarray}\label{YPI}
Y({\bf k})=-\int_{-\infty}^\infty\frac{d\omega}{\pi}
\frac{\omega |\Delta(\omega,{\bf k})|}{\left[\omega^2+{\cal W}^2(\omega,{\bf k})\right]^2}\frac{\partial |\Delta(\omega,{\bf k})|}{\partial \omega}.
\end{eqnarray}
This expression shows obviously that $\rho_n$ is generically nonzero with a frequency-dependent gap function. For realistic boson-mediated interaction, $|\Delta(\omega,{\bf k})|$ decreases monotonically
with increasing $|\omega|$ (see Appendix B). Therefore, $Y({\bf k})$ is normally positive, leading to a nonvanishing  $\rho_n$ and hence a quantum reduction of $\rho_s$.  One the other hand, we expect that 
the finite-temperature behavior of the superfluid density $\rho_s$ is regular. It drops down with increasing temperature and finally approaches zero at the superfluid transition temperature.

The result here based on the mean-field theory captures of the essential physics of the DBE. Inclusion of the quantum fluctuations does not change the result qualitatively. Actually, the quantum fluctuations
bring a correction to $\rho_n$ which should not be negative \cite{Taylor2006,Brendan2017}. Even though here we assume a spin-singlet pairing, this generic result does not rely on the pairing symmetry, e.g., 
it also applies to the $p$-wave topological superfluid in a single-component 2D Fermi gas \cite{Wu2016}, with interaction mediated by a 3D Bose-Einstein condensate.

However, a puzzle may appear if we consider a spin-$1/2$ Fermi gas with balanced spin populations with static $s$-wave two-body interaction between the unlike spins. Such a system has Galilean invariance and hence the superfluid density should equal the total fermion density. However, a puzzle appears if we consider dressing the interaction by using the random phase approximation (RPA), or consider the so-called induced interaction \cite{Bulgac2009}. In this case, the PRA improved interaction is no longer static and hence the superfluid density may be reduced according to our formula (\ref{YPI}).  To solve this puzzle, we recall that the superfluid density is defined as a linear response to an infinitesimal uniform superfluid flow. On the other hand, the RPA improved two-body interaction is built by using the fermion Green's function. In the presence of a superfluid flow, the fermion Green's function is given by Eq. (\ref{GA}).  If we calculate the superfluid density with the RPA improved two-body interaction, we also need to consider the modification of the fermion Green's function due to the superfluid flow. Therefore, there should be an additional contribution to the superfluid density from the RPA improved two-body interaction. If the RPA theory is self-consistent and compatible with the Galilean invariance, the additional contribution from the RPA should compensate the reduction due to the dynamical boson exchange effect. The total superfluid density thus remains the total fermion density.

Next we consider the spin response. We introduce a Zeeman term ${\cal S}_h=h\left[\rho_\uparrow(0)-\rho_\downarrow(0)\right]$ in the effective action, with $h$ being the Zeeman field. 
For an infinitesimal $h$, the change in the grand potential reads $\Omega(h)-\Omega(0)=-\frac{1}{2}\chi h^2+O(h^4)$ and the induced magnetization is given by $n_\uparrow-n_\downarrow=\chi h+O(h^3)$. 
The spin susceptibility $\chi$ can be expressed as
\begin{eqnarray}
\chi=\frac{2}{\cal V}\sum_{\bf k}Y({\bf k}),
\end{eqnarray}
where $Y({\bf k})$ is the same function defined in (\ref{Yfunction}). Without DBE, it is obvious that $\chi=0$ at $T=0$. The DBE thus leads to a nonzero $\chi$. In this case, applying a finite Zeeman field $h$ will induce a finite magnetization even though $h$ is below the Pauli limit (see Appendix E). 

While the DBE generically leads to nonvanishing normal fluid density and spin susceptibility, a natural question is whether they are sizable and hence can be probed experimentally.  
From the gap equation (\ref{GEQ}),  we can derive a useful expression for $\partial |\Delta|/\partial\omega$ (see Appendix B): 
\begin{eqnarray}
\frac{\partial|\Delta(\omega,{\bf k})|}{\partial\omega}=-\int_{\nu,{\bf p}}\frac{\partial U(K-P)}{\partial\omega}\frac{|\Delta(\nu,{\bf p})|}{\nu^2+{\cal W}^2(\nu,{\bf p})}.
\end{eqnarray}
Together with Eq. (\ref{YPI}),
we expect that sizable values of $\rho_n$ and $\chi$ require a large magnitude of the gap function, i.e., strong pairing. Therefore, in the weak pairing limit, $|\Delta|\ll\varepsilon_{\rm F}$, 
$\rho_n$ and $\chi$ become vanishingly small.  On the other hand, for very strong pairing, the system becomes a Bose condensate of bound pairs. In this regime, we have $\mu_{\rm f}<0$ and 
$|\mu_{\rm f}|\gg|\Delta|$, indicating that $\rho_n$ and $\chi$ are also rather small. 

One should not conflate the superfluid density with the condensate density. The condensate number of fermion pairs can be evaluated by using its definition $N_0=\int d{\bf r}\int d{\bf r}^\prime
|\psi_\downarrow({\bf r})\psi_\uparrow({\bf r}^\prime)|^2$. In the mean-field theory, the condensate density $n_0$ at $T=0$ reads
\begin{eqnarray}
n_0=\frac{1}{{\cal V}}\sum_{\bf k}\left[\int_{-\infty}^\infty\frac{d\omega}{2\pi}
\frac{ |\Delta(\omega,{\bf k})|}{\omega^2+{\cal W}^2(\omega,{\bf k})}\right]^2.
\end{eqnarray}
The behavior of $n_0$ is thus quite different from the superfluid density $\rho_s$. In the weak pairing limit, we have $n_0\rightarrow 0$ while $\rho_s\rightarrow n_{\rm f}$. With increasing pairing strength, the condensate density gets enhanced but the superfluid density is suppressed due to the DBE, leading to opposite behavior of condensation and superfluidity.

\section{Single-component superfluid: A toy Yukawa model}

We first consider a system in which only the fermion component is a superfluid and the bosons play the role of force carriers. We study a toy model, fermions attracting each other via a real scalar mode $\phi$, of which the Lagrangian density  reads
\begin{eqnarray}
{\cal L}_{\rm b}=\frac{1}{2}\left[(\partial_\tau\phi)^2+v_\phi^2(\nabla\phi)^2+m_\phi^2\phi^2\right].
\end{eqnarray}
The fermions and bosons interact via a Yukawa coupling ${\cal L}_{\rm bf}=g\phi\sum_\sigma\psi^\dagger_\sigma\psi^{\phantom{\dag}}_\sigma$.
Integrating out the scalar mode, we obtain the pairing interaction
\begin{eqnarray}
U(q_l,{\bf q})=-\frac{g^2}{q_l^2+v_\phi^2{\bf q}^2+m_\phi^2}.
\end{eqnarray}
The velocity parameter $v_\phi$ \cite{Yukawa-note} and the mass parameter $m_\phi$ control the interaction retardation and also the pairing strength. In the static approximation, i.e., discarding $q_l^2$, the
pairing interaction reduces to a static Yukawa-type potential. 

For $s$-wave pairing, the gap function depends only on $k=|{\bf k}|$ and can be set to be real. We solve the gap and number equations for $s$-wave pairing at $T=0$ (see Appendixes B and C). 
In  Fig. \ref{fig1}(a), we show the the dependence of $\rho_n/n_{\rm f}$ and $\chi/\chi_0$ on the velocity parameter $v_\phi$, where $\chi_0=3n_{\rm f}/(2\varepsilon_{\rm F})$ is the spin susceptibility of a 
noninteracting Fermi gas.  With decreasing values of $v_\phi/v_{\rm F}$, the interaction retardation becomes more pronounced and hence $\rho_n$ and $\chi$ become enhanced. Figure \ref{fig1}(b) shows the 
dependence of $\rho_n$ and $\chi$ on the mass parameter $m_\phi$ at vanishingly small velocity $v_\phi/v_{\rm F}\rightarrow0$. In this case, the gap function depends only on the frequency $\omega$. 
With increasing values of $m_\phi$, the pairing becomes weaker [Fig. \ref{fig1}(c)], and hence the derivative $\partial\Delta/\partial\omega$ gets smaller. Accordingly, we find that $\rho_n$ and $\chi$ become vanishingly small at large $m_\phi$.

In a single fermion superfluid, the speed of the second sound is given by \cite{Taylor2009,Salasnich2010}
 \begin{eqnarray}
c_2=\sqrt{\frac{1}{m_{\rm f}}\frac{s^2}{(\partial s/\partial T)_{n_{\rm f}}}\frac{\rho_s}{\rho_n}}
\end{eqnarray}
In the absence of DBE, $c_2$ approaches a nonzero constant when $T\rightarrow0$ \cite{Taylor2009,Salasnich2010}, because of the cooperative low-$T$ behavior of the entropy per fermion $s$ and $\rho_n$. However, the DBE leads to $c_2\sim T^2$ for $T\rightarrow0$ due to the nonvanishing $\rho_n$ at $T=0$.  

\begin{figure}
\centering{}\includegraphics[width=0.45\textwidth]{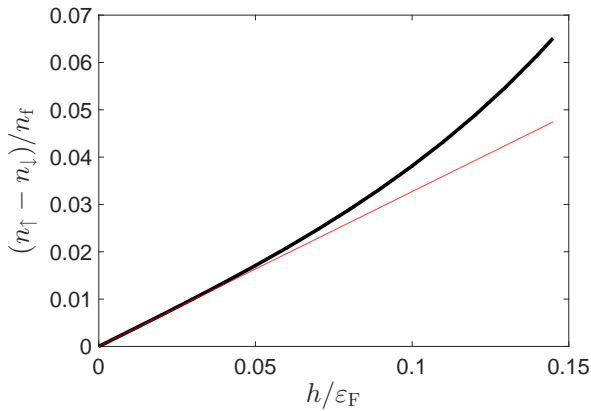} \caption{ The spin imbalance $(n_\uparrow-n_\downarrow)/n_{\rm f}$ as a function of the Zeeman field $h$ (scaled by $\varepsilon_{\rm F}$) 
for $v_\phi/v_{\rm F}=0$ and $m_\phi/\varepsilon_{\rm F}=0.1$ in the toy Yukawa model. The Pauli limit is located at $h\simeq0.145\varepsilon_{\rm F}$. The thin red line denotes the linear approximation
$n_\uparrow-n_\downarrow\simeq\chi h$.
\label{fig2}}
\end{figure}


Next we study the system at finite Zeeman field $h$ at $T=0$  (see Appendix E). With increasing $h$, the superfluid state persists up to the Pauli limit determined by the sign change of the grand potential difference $\Omega_{\rm S}-\Omega_{\rm N}$. In Fig. \ref{fig2}, we show that the spin imbalance is nonzero in the superfluid phase. At small $h$, the spin imbalance is well given by $n_\uparrow-n_\downarrow\simeq\chi h$. Thus the nonzero spin imbalance generated in the superfluid phase is caused by the DBE. For static pairing interaction, the spin imbalance keeps vanishing in the superfluid phase below the Pauli limit.

\section{Double superfluidity: Fermi-Bose mixtures}

A possible cold atom system to test our predictions is a dilute Fermi-Bose mixture which has been realized in recent cold atom experiments 
\cite{BF-Review, BF-INT01,BF-INT02,BF-Mix01,BF-Mix02,BF-Mix03,BF-Mix04,BF-Mix05,BF-Mix06}.  We consider a mixture composed of two-component fermions and weakly interacting bosons 
of mass $m_{\rm b}$.  With a boson-fermion interaction, an effective interaction between fermions can be induced by the Bogoliubov phonon mode of the BEC \cite{Heiselberg2000,Viverit2000,Stoof2000}. The bosons are described by a complex scalar field $\phi$ and its Lagrangian density reads
\begin{eqnarray}
{\cal L}_{\rm b}=\phi^\dagger\left(-\partial_\tau+\frac{\nabla^2}{2m_{\rm b}}\right)\phi-\frac{2\pi a_{\rm bb}}{m_{\rm b}}|\phi|^4,
\end{eqnarray}
where $a_{\rm bb}$ is the boson-boson scattering length. The boson-fermion interaction is given by
\begin{eqnarray}
{\cal L}_{\rm bf}=g_{\rm bf}|\phi|^2\sum_\sigma\psi^\dagger_\sigma\psi^{\phantom{\dag}}_\sigma,
\end{eqnarray}
where $g_{\rm bf}=4\pi a_{\rm bf}/m_{\rm bf}$ 
is the boson-fermion coupling, with the boson-fermion scattering length $a_{\rm bf}$ and the reduced mass $m_{\rm bf}=2m_{\rm b}m_{\rm f}/(m_{\rm b}+m_{\rm f})$. 

Within the Bogoliubov theory, the action for the BEC reads ${\cal S}_{\rm b}=\sum_Q[-iq_l+E_{\rm b}({\bf q})]\varphi^*(Q)\varphi(Q)$, where $\varphi(Q)$ is phonon field and 
$E_{\rm b}({\bf q})=\sqrt{v_{\rm B}^2{\bf q}^2+\varepsilon^2_{\rm b}({\bf q})}$ is the Bogoliubov spectrum, with $\varepsilon_{\rm b}({\bf q})={\bf q}^2/(2m_{\rm b})$ and 
the phonon velocity $v_{\rm B}=\sqrt{4\pi a_{\rm bb}n_{\rm b}/m_{\rm b}^2}$. Here the boson condensate density $n_{\rm b}$ equals the boson number density at $T=0$.  
In terms of the phonon field,
the boson-fermion interaction can be expressed as ${\cal S}_{\rm bf}={\cal V}^{-1/2}\sum_\sigma\sum_Q M({\bf q})[\varphi^*(Q)+\varphi(-Q)]\rho_\sigma(Q)$, where 
$M({\bf q})=g_{\rm bf}\sqrt{n_{\rm b}\varepsilon_{\rm b}({\bf q})/E_{\rm b}({\bf q})}$.  Integrating out the phonon mode, 
we obtain the pairing interaction $U(q_l,{\bf q})=u_{\rm ff}({\bf q})+U_{\rm ind}(q_l,{\bf q})$, where $u_{\rm ff}({\bf q})$ is the direct instantaneous interaction. 
The induced interaction mediated by the Bogoliubov phonon is given by \cite{Heiselberg2000,Viverit2000,Stoof2000}
\begin{eqnarray}\label{Ufb}
U_{\rm ind}(q_l,{\bf q})=-\frac{2g_{\rm bf}^2n_{\rm b}\varepsilon_{\rm b}({\bf q})}{q_l^2+[E_{\rm b}({\bf q})]^2}.
\end{eqnarray}

While the Galilean invariance of the fermion sector is violated due to the phonon exchange, the whole system is Galilean invariant. We note that the boson component is also a superfluid and hence this double-superfluid system should be described by the three-fluid theory \cite{Andreev1975, Three-Fluid}.  We thus introduce two superfluid flow velocities, 
${\bf v}_{\rm f}$ and ${\bf v}_{\rm b}$, for the fermion and boson parts respectively. The change in the free energy is now given by \cite{Note-Dij}
\begin{eqnarray}
F({\bf v}_{\rm f},{\bf v}_{\rm b})-F({\bf 0},{\bf 0})=\frac{1}{2}\sum_{\rm i,j=f,b}D_{\rm ij}{\bf v}_{\rm i}\cdot{\bf v}_{\rm j}+O({\bf v}_{\rm i}^4).
\end{eqnarray}
The diagonal term for the fermionic part has been calculated, $D_{\rm ff}=m_{\rm f}(n_{\rm f}-\rho_{\rm d})$, where the reduction $\rho_{\rm d}$ is just given by 
the expression of $\rho_n$ in Eq. (\ref{RHOn}).  The off-diagonal terms, $D_{\rm fb}=D_{\rm bf}$, represents the Andreev-Bashkin drag effect \cite{Andreev1975}.  The Galilean invariance of the whole system requires $D_{\rm ff}+D_{\rm fb}=m_{\rm f}n_{\rm f}$ \cite{Nespolo2017}. Thus we obtain 
\begin{eqnarray}\label{ABdrag}
D_{\rm fb}=D_{\rm bf}=m_{\rm f}\rho_{\rm d}.
\end{eqnarray}
Therefore, the quantum reduction $\rho_{\rm d}$ of the fermion sector just gives rise to the Andreev-Bashkin drag effect. To prove Eq. (\ref{ABdrag}), we note that the bosonic flow ${\bf v}_{\rm b}$ leads to a shift $iq_l\rightarrow iq_l-{\bf q}\cdot {\bf v}_{\rm b}$ 
in the induced interaction (\ref{Ufb}), which further induces a change in the gap function $\Delta(K)$. To the leading order in ${\bf v}_{\rm b}$, the change at $T=0$ reads 
$i{\bf k}\cdot{\bf v}_{\rm b}\partial\Delta/\partial\omega$.  Collecting the terms proportional to ${\bf v}_{\rm f}\cdot{\bf v}_{\rm b}$ and using Eq. (\ref{YPI}), we find that the off-diagonal coefficients  
just equal $m_{\rm f}\rho_{\rm d}$. This proof also shows that the mean-field theory for the fermionic superfluidity is compatible with the Galilean invariance and hence is qualitatively reliable for the predictions 
of the superfluid and normal densities. 

\begin{figure}
\centering{}\includegraphics[width=0.53\textwidth]{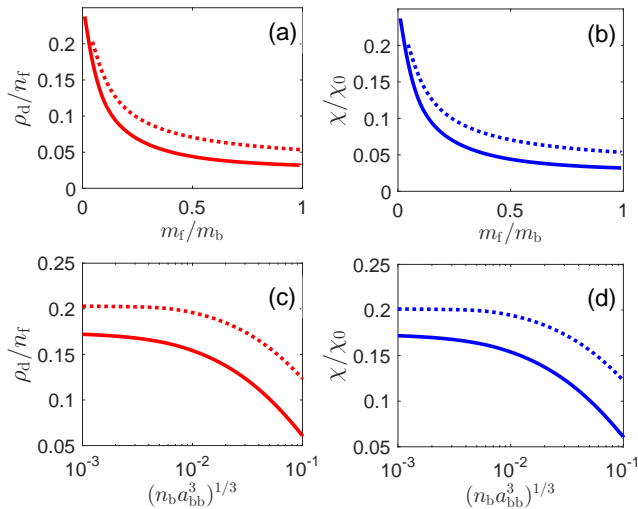} \caption{Results for the Fermi-Bose mixture. (a)(b) The dependence of $\rho_{\rm d}$ and $\chi$ 
on the mass ratio $m_{\rm f}/m_{\rm b}$ at $(n_{\rm b}a_{\rm bb}^3)^{1/3}=0.001$. 
(c)(d) The dependence of $\rho_{\rm d}$ and $\chi$ on the boson-boson interaction parameter $(n_{\rm b}a_{\rm bb}^3)^{1/3}$ at $m_{\rm f}/m_{\rm b}=6/133$ ($^6$Li-$^{133}$Cs mixture).
The boson-fermion coupling is chosen as $(n_{\rm b}a_{\rm bf}^3)^{1/3}=0.1$ (solid lines) and $0.15$ (dotted lines). The fermion density is $n_{\rm f}=0.1n_{\rm b}$.
\label{fig3}}
\end{figure}


We solve the gap and number equations for $s$-wave pairing for $u_{\rm ff}=0$ at $T=0$  (see Appendixes B and C). The result depends on the interaction parameters 
$(n_{\rm b}a_{\rm bb}^3)^{1/3}$ and $(n_{\rm b}a_{\rm bf}^3)^{1/3}$,  the density ratio $n_{\rm f}/n_{\rm b}$, and the mass ratio $m_{\rm f}/m_{\rm b}$.  The interaction retardation can be characterized by the velocity ratio
\begin{eqnarray}
\frac{v_{\rm B}}{v_{\rm F}}=\frac{2\sqrt{\pi}}{(3\pi^2)^{1/3}}\frac{m_{\rm f}}{m_{\rm b}}\frac{(n_{\rm b}a_{\rm bb}^3)^{1/6}}{(n_{\rm f}/n_{\rm b})^{1/3}}.
\end{eqnarray}
For small $v_{\rm B}/v_{\rm F}$, we expect that the retardation is significant and hence $\rho_{\rm d}$ and $\chi$ reach sizable values. In Fig. \ref{fig3} we show the results of $\rho_{\rm d}$ and $\chi$ for 
$n_{\rm f}/n_{\rm b}=0.1$ and two values of the boson-fermion coupling $(n_{\rm b}a_{\rm bf}^3)^{1/3} (0.1$ and $0.15$).  From Figs. \ref{fig3} (a) and 3(b), we find that a small mass ratio 
$m_{\rm f}/m_{\rm b}$ enhances $\rho_{\rm d}$ and $\chi$ due to strong IRE. Figures. \ref{fig3} (c) and 3(d) show that a smaller boson-boson interaction parameter $(n_{\rm b}a_{\rm bb}^3)^{1/3}$ 
leads to larger $\rho_{\rm d}$ and $\chi$, also consistent with the interaction retardation picture. 

We expect that the recently realized $^6$Li-$^{174}$Yb \cite{BF-Mix05} and $^6$Li-$^{133}$Cs \cite{BF-Mix06} mixtures are good candidates to 
test our theoretical predictions and explore the Andreev-Bashkin drag effect. Our simple calculation shows that $\rho_{\rm d}/n_{\rm f}$ can reach the order $10\%$ in the $^6$Li-$^{133}$Cs mixture \cite{ABF-note}. Note that here we have neglected the direct pairing interaction $u_{\rm ff}$ which can be tuned by using the method of Feshbach resonance \cite{FR-RMP}. Turning on this resonant interaction will greatly enhance the magnitude of the gap function $\Delta$ and hence the derivative $\partial\Delta/\partial\omega$  (see Appendix D), leading to a much stronger coupling between the two superfluid components.   On the other hand, for realistic experimental systems, we need to consider the trapping potential. In this case, we can use the local density approximation to calculate the local superfluid density or the superfluid density profile. Within the local density approximation, our theoretical predictions are still valid for the local superfluid density. The total superfluid fraction can also be computed and could be compared with the future experimental measurements.

\section{Summary}

We have shown that the interaction retardation due to DBE in a fermionic superfluid leads to a quantum reduction of the superfluid density.  For $s$-wave pairing, it also leads to a nonzero spin susceptibility at $T=0$ and hence allows spin-imbalanced pairing when a Zeeman field is applied.  For the double-superfluid system in ultracold Fermi-Bose mixtures,  the quantum reduction also gives rise to the Andreev-Bashkin drag effect. These effects become sizable at strong pairing and hence can be probed in cold atom experiments. The superfluid or normal fraction can be probed by measuring the moment of inertia in the presence of a slow rotation \cite{Stringari1999,Foot2000,Grimm2011,Baym2013,Cooper2010} and can be extracted by measuring the second sound \cite{Tey2013}. On the other hand, the superfluid  fraction can be computed by using the quantum Monte Carlo method \cite{Ceperly1987,Akkineni2007}. 

Other possible candidate systems to explore the effect of DBE on fermion superfluidity may include an ultracold Fermi gas with an effective interaction mediated by the cavity modes \cite{Guo2012} and the exciton-polariton mediated superconductivity in two-dimensional electron gases \cite{Laussy2010}. 
Since the nuclear force includes contributions from meson exchange, the DBE may also provide a new mechanism to reduce the superfluid density of neutron matter, which is crucial for models of neutron star glitches based on neutron superfluidity \cite{Watanabe2017,Chamel2012}.

\begin{acknowledgments}
We thank Prof. Hui Hu for a critical reading of the manuscript and helpful suggestions,  and Dr. Zhigang Wu for useful discussions. The work was supported
by the National Natural Science Foundation of China (Grant No. 11775123) and the National Key R\&D Program of China (Grant No. 2018YFA0306503). 
\end{acknowledgments}

\appendix

\begin{widetext}

\section{Trick of integration by parts}
We prove Eq. (13) by using a simple trick of integration by parts.  We consider the integral
\begin{eqnarray}
Y=\int_{-\infty}^\infty\frac{d\omega}{2\pi}\frac{\omega^2-\xi^2-|\Delta(\omega)|^2}{\left[\omega^2+\xi^2+|\Delta(\omega)|^2\right]^2},
\end{eqnarray}
which can be written in an alternative form
\begin{eqnarray}
Y=\int_{-\infty}^\infty\frac{d\omega}{2\pi}\left\{\frac{2\omega^2}{\left[\omega^2+\xi^2+|\Delta(\omega)|^2\right]^2}-\frac{1}{\omega^2+\xi^2+|\Delta(\omega)|^2}\right\}.
\end{eqnarray}
Using the identity
\begin{eqnarray}
\frac{2\omega^2}{\left[\omega^2+\xi^2+|\Delta(\omega)|^2\right]^2}=-\omega\frac{\partial}{\partial\omega}\frac{1}{\omega^2+\xi^2+|\Delta(\omega)|^2}
-\frac{\omega}{\left[\omega^2+\xi^2+|\Delta(\omega)|^2\right]^2}\frac{\partial|\Delta(\omega)|^2}{\partial\omega}
\end{eqnarray}
and performing integration by parts, we obtain
\begin{eqnarray}
Y=-\int_{-\infty}^\infty\frac{d\omega}{\pi}\frac{\omega|\Delta(\omega)|}{\left[\omega^2+\xi^2+|\Delta(\omega)|^2\right]^2}\frac{\partial |\Delta(\omega)|}{\partial\omega}.
\end{eqnarray}
Note that in this proof, we do not require $|\Delta(\omega)|\rightarrow 0$ for $\omega\rightarrow\pm\infty$.

\section{Gap equation for $s$-wave pairing}

For $s$-wave pairing, the gap function depends only on $k=|{\bf k}|$ and can be set to be real without loss of generality. The $s$-wave gap equation can be expressed as
\begin{eqnarray}\label{gapeq}
\Delta(\nu,p)=\int_0^\infty\frac{d\omega}{\pi}\int_0^\infty\frac{k^2dk}{2\pi^2}{\cal K}_s(\nu,p;\omega,k)\frac{\Delta(\omega,k)}{\omega^2+\xi_k^2+\Delta^2(\omega,k)},
\end{eqnarray}
where we have used the fact $\Delta(\omega,k)=\Delta(-\omega,k)$. The $s$-wave kernel function ${\cal K}_s(\nu,p;\omega,k)$ is given by
\begin{eqnarray}
{\cal K}_s(\nu,p;\omega,k)=\frac{1}{2}\left[U_s(\nu-\omega;p,k)+U_s(\nu+\omega;p,k)\right].
\end{eqnarray}
Here $U_s(\nu-\omega;p,k)$ is the $s$-wave interaction obtained by performing the angle integration,
\begin{eqnarray}
U_s(\nu-\omega;p,k)=-\frac{1}{2}\int_{0}^\pi \sin\theta d\theta U(\nu-\omega,{\bf p}-{\bf k}),
\end{eqnarray}
where $\theta$ is the angle between ${\bf p}$ and ${\bf k}$. For the Yukawa model, we have
\begin{eqnarray}
U_s(\nu-\omega;p,k)=\frac{g^2}{4v_\phi^2pk}
\ln\left[\frac{(\nu-\omega)^2+v_\phi^2(p+k)^2+m_\phi^2}{(\nu-\omega)^2+v_\phi^2(p-k)^2+m_\phi^2}\right].
\end{eqnarray}
For the dilute Fermi-Bose mixture with $u_{\rm ff}=0$,  $U_s(\nu-\omega;p,k)$ is given by
\begin{eqnarray}
U_s(\nu-\omega;p,k)&=&\frac{g_{\rm bf}^2n_{\rm b}m_{\rm b}}{2pk}\Bigg\{
\ln\left[\frac{(\nu-\omega)^2+2g_{\rm bb}n_{\rm b}\frac{(p+k)^2}{2m_{\rm b}}+\frac{(p+k)^4}{4m_{\rm b}^2}}{(\nu-\omega)^2+2g_{\rm bb}n_{\rm b}\frac{(p-k)^2}{2m_{\rm b}}+\frac{(p-k)^4}{4m_{\rm b}^2}}\right]\nonumber\\
&+&\frac{2g_{\rm bb}n_{\rm b}}{\sqrt{(\nu-\omega)^2-(g_{\rm bb}n_{\rm b})^2}}
\left[\arctan\frac{g_{\rm bb}n_{\rm b}+\frac{(p-k)^2}{2m_{\rm b}}}{\sqrt{(\nu-\omega)^2-(g_{\rm bb}n_{\rm b})^2}}
-\arctan\frac{g_{\rm bb}n_{\rm b}+\frac{(p+k)^2}{2m_{\rm b}}}{\sqrt{(\nu-\omega)^2-(g_{\rm bb}n_{\rm b})^2}}\right]\Bigg\}.
\end{eqnarray}
Note that for $(\nu-\omega)^2<(g_{\rm bb}n_{\rm b})^2$,  the inverse tangent function is understood as $\arctan(ix)=\ln[(1+x)/(1-x)]/(2i)$.  

On the other hand, from the gap equation (\ref{gapeq}), we obtain
\begin{eqnarray}
\frac{\partial\Delta(\omega,k)}{\partial\omega}=\int_0^\infty\frac{d\nu}{\pi}\int_0^\infty\frac{p^2dp}{2\pi^2}\frac{\partial {\cal K}_s(\omega,k;\nu,p)}{\partial\omega}
\frac{\Delta(\nu,p)}{\nu^2+\xi_p^2+\Delta^2(\nu,p)},
\end{eqnarray}
Thus we estimate that the magnitude of the derivative $\partial\Delta/\partial\omega$ depends on two components: the derivative $\partial {\cal K}_s/\partial\omega$ representing the DBE 
effect and the magnitude of the gap function. For the Yukawa model, we have
\begin{eqnarray}
\frac{\partial {\cal K}_s(\omega,k;\nu,p)}{\partial\omega}=-g^2\left\{\frac{\omega-\nu}{[(\omega-\nu)^2+v_\phi^2(k^2+p^2)+m_\phi^2]^2+4v_\phi^4k^2p^2}+(\nu\rightarrow-\nu)\right\}.
\end{eqnarray}
Thus $\partial\Delta/\partial\omega$ is zero at $\omega=0$ and is negative for $\omega>0$.

Using the dimensionless quantities,
\begin{eqnarray}
x=\frac{\omega}{\varepsilon_{\rm F}}, \ \ \ \ \ \  x^\prime=\frac{\nu}{\varepsilon_{\rm F}},\ \ \ \ \ \ y=\frac{k}{k_{\rm F}}, \ \ \ \ \ \  y^\prime=\frac{p}{k_{\rm F}},\ \ \ \ \ \ 
\tilde{\Delta}=\frac{\Delta}{\varepsilon_{\rm F}}, \ \ \ \ \ \  \tilde{\mu}=\frac{\mu}{\varepsilon_{\rm F}},\
\end{eqnarray}
we can express the gap equation as
\begin{eqnarray}
\tilde{\Delta}(x^\prime,y^\prime)=\int_0^\infty dx\int_0^\infty y^2dy \tilde{\cal K}_s(x^\prime,y^\prime;x,y)
\frac{\tilde{\Delta}(x,y)}{x^2+(y^2-\tilde{\mu})^2+\tilde{\Delta}^2(x,y)}.
\end{eqnarray}
For the Yukawa model, the dimensionless kernel function $\tilde{\cal K}_s(x^\prime,y^\prime;x,y)$ is given by
\begin{eqnarray}
\tilde{\cal K}_s(x^\prime,y^\prime;x,y)&=&\frac{g^2}{8\pi^3 v_{\rm F}^3}\frac{1}{\gamma^2 yy^\prime}
\ln\left[\frac{(x^\prime-x)^2+4\gamma^2(y+y^\prime)^2+\tilde{m}_\phi^2}{(x^\prime-x)^2+4\gamma^2(y-y^\prime)^2+\tilde{m}_\phi^2}\right]+(x\rightarrow-x),
\end{eqnarray}
where $\gamma=v_\phi/v_{\rm F}$ and $\tilde{m}_\phi=m_\phi/\varepsilon_{\rm F}$. For the Fermi-Bose mixture with $u_{\rm ff}=0$, we have
\begin{eqnarray}
\tilde{\cal K}_s(x^\prime,y^\prime;x,y)&=&\frac{(1+\alpha)^2}{\alpha \pi yy^\prime}\lambda_1\Bigg\{
\ln\left[\frac{(x^\prime-x)^2+2\alpha^2\lambda_2(y+y^\prime)^2+\alpha^2(y+y^\prime)^4}{(x^\prime-x)^2+2\alpha^2\lambda_2(y-y^\prime)^2+\alpha^2(y-y^\prime)^4}\right]\nonumber\\
&&+\frac{\alpha\lambda_2}{\sqrt{(x^\prime-x)^2-\alpha^2\lambda_2^2}}
\left[\arctan\frac{\alpha\lambda_2+\alpha(y-y^\prime)^2}{\sqrt{(x^\prime-x)^2-\alpha^2\lambda_2^2}}
-\arctan\frac{\alpha\lambda_2+\alpha(y+y^\prime)^2}{\sqrt{(x^\prime-x)^2-\alpha^2\lambda_2^2}}\right]\Bigg\}\nonumber\\
&&+(x\rightarrow-x),
\end{eqnarray}
where
\begin{eqnarray}
\alpha=\frac{m_{\rm f}}{m_{\rm b}},\ \ \ \ \  \lambda_1=\frac{(n_{\rm b}a_{\rm bf}^3)^{2/3}}{\left(3\pi^2n_{\rm f}/n_{\rm b}\right)^{1/3}},\ \ \ \ \ 
\lambda_2=8\pi \frac{(n_{\rm b}a_{\rm bb}^3)^{1/3}}{(3\pi^2n_{\rm f}/n_{\rm b})^{2/3}}.
\end{eqnarray}
The solution thus depends on four quantities: $(n_{\rm b}a_{\rm bf}^3)^{1/3}$,  $(n_{\rm b}a_{\rm bb}^3)^{1/3}$, $\alpha$, and $n_{\rm f}/n_{\rm b}$.

Meanwhile, the number equation becomes
\begin{eqnarray}
\int_0^\infty y^2dy \left[1-\frac{2}{\pi}\int_0^\infty dx\frac{y^2-\tilde{\mu}}{x^2+(y^2-\tilde{\mu})^2+\tilde{\Delta}^2(x,y)}\right]=\frac{2}{3}.
\end{eqnarray}
The normal density $\rho_n$ and the spin  susceptibility $\chi$ can be expressed as
\begin{eqnarray}
\frac{\rho_n}{n_{\rm f}}&=&-\frac{4}{\pi}\int_0^\infty xdx\int_0^\infty y^4dy \frac{\tilde{\Delta}(x,y)}{\left[x^2+(y^2-\tilde{\mu})^2+\tilde{\Delta}^2(x,y)\right]^2}\frac{\partial\tilde{\Delta}(x,y)}{\partial x},\nonumber\\
\frac{\chi}{\chi_0}&=&-\frac{4}{\pi}\int_0^\infty xdx\int_0^\infty y^2dy \frac{\tilde{\Delta}(x,y)}{\left[x^2+(y^2-\tilde{\mu})^2+\tilde{\Delta}^2(x,y)\right]^2}\frac{\partial\tilde{\Delta}(x,y)}{\partial x}.
\end{eqnarray}

\section{Treatment of the number equation}

For balanced spin populations, the number equation can be expressed as $n_{\rm f}=(2/{\cal V})\sum_{\bf k}n_{\bf k}$, where
the fermion momentum distribution $n_{\bf k}$ is formally given by
\begin{eqnarray}
n_{\bf k}=-\frac{1}{\beta}\sum_{n}\frac{(i\omega_n+\xi_{\bf k})e^{i\omega_n0^+}}{\omega_n^2+\xi_{\bf k}^2+|\Delta(\omega_n,{\bf k})|^2}.
\end{eqnarray}
The convergent factor $e^{i\omega_n0^+}$ is not convenient for a numerical calculation.  Here we provide a useful treatment at $T=0$ and it can be easily generalized to $T\neq0$.
At $T=0$, we have
\begin{eqnarray}
n_{\bf k}=-\int_{-\infty}^\infty\frac{d\omega}{2\pi}\frac{(i\omega+\xi_{\bf k})e^{i\omega0^+}}{\omega^2+\xi_{\bf k}^2+|\Delta(\omega,{\bf k})|^2}.
\end{eqnarray}
Noting that the gap function becomes vanishingly small at large $|\omega|$, we introduce a large cutoff $\omega_c$ and divide the integration into three parts,
\begin{eqnarray}
n_{\bf k}=I_1+I_2+I_3,
\end{eqnarray}
where
\begin{eqnarray}
&&I_1=-\int_{-\omega_c}^{\omega_c}\frac{d\omega}{2\pi}\frac{(i\omega+\xi_{\bf k})e^{i\omega0^+}}{\omega^2+\xi_{\bf k}^2+|\Delta(\omega,{\bf k})|^2},\nonumber\\
&&I_2=-\int_{\omega_c}^{\infty}\frac{d\omega}{2\pi}\frac{(i\omega+\xi_{\bf k})e^{i\omega0^+}}{\omega^2+\xi_{\bf k}^2+|\Delta(\omega,{\bf k})|^2},\nonumber\\
&&I_3=-\int_{-\infty}^{-\omega_c}\frac{d\omega}{2\pi}\frac{(i\omega+\xi_{\bf k})e^{i\omega0^+}}{\omega^2+\xi_{\bf k}^2+|\Delta(\omega,{\bf k})|^2}.
\end{eqnarray}
The convergent factor only guarantees the convergence for $|\omega|\rightarrow\infty$. Therefore, we can get rid of it in $I_1$ and obtain
\begin{eqnarray}
I_1=-\int_{0}^{\omega_c}\frac{d\omega}{\pi}\frac{\xi_{\bf k}}{\omega^2+\xi_{\bf k}^2+|\Delta(\omega,{\bf k})|^2}.
\end{eqnarray}
Here we have use the fact $\Delta(\omega,{\bf k})=\Delta(-\omega,{\bf k})$. Since the cutoff $\omega_c$ is large, we can neglect $\Delta(\omega,{\bf k})$ in $I_2$ and $I_3$. It becomes exact
when we set $\omega_c\rightarrow\infty$ finally. Thus we have
\begin{eqnarray}
I_2\simeq\int_{\omega_c}^{\infty}\frac{d\omega}{2\pi}\frac{e^{i\omega0^+}}{i\omega-\xi_{\bf k}}, \ \ \ \ \ I_3\simeq\int_{-\infty}^{-\omega_c}\frac{d\omega}{2\pi}\frac{e^{i\omega0^+}}{i\omega-\xi_{\bf k}}.
\end{eqnarray}
They can be evaluated by using a contour integration. We obtain
\begin{eqnarray}
I_2=-\int_0^{\infty}\frac{idy}{2\pi}\frac{e^{-y0^+}}{y+\xi_{\bf k}-i\omega_c}, \ \ \ \ \ I_3=\int_0^{\infty}\frac{idy}{2\pi}\frac{e^{-y0^+}}{y+\xi_{\bf k}+i\omega_c}.
\end{eqnarray}
The convergent factor can be dropped when summing $I_2$ and $I_3$. We obtain
\begin{eqnarray}
I_2+I_3=\frac{\omega_c}{\pi}\int_0^{\infty}\frac{dy}{(y+\xi_{\bf k})^2+\omega_c^2}=\frac{1}{2}-\frac{1}{\pi}\arctan\frac{\xi_{\bf k}}{\omega_c}.
\end{eqnarray}
Finally, setting $\omega_c\rightarrow\infty$, we obtain
\begin{eqnarray}
n_{\bf k}=\frac{1}{2}-\int_0^\infty\frac{d\omega}{\pi}\frac{\xi_{\bf k}}{\omega^2+\xi_{\bf k}^2+|\Delta(\omega,{\bf k})|^2}.
\end{eqnarray}
If the gap function depends only on the momentum, i.e., $\Delta(\omega,{\bf k})=\Delta_{\bf k}$, we recover the known result
\begin{eqnarray}
n_{\bf k}=\frac{1}{2}\left(1-\frac{\xi_{\bf k}}{\sqrt{\xi_{\bf k}^2+|\Delta_{\bf k}|^2}}\right).
\end{eqnarray}

\section{Effect of an instantaneous pairing interaction}

With an additional instantaneous pairing interaction, the gap equation reads
\begin{eqnarray}
\Delta(P)=-\frac{1}{\beta {\cal V}}\sum_{K}\left[u_{\rm ff}({\bf p}-{\bf k})+U_{\rm ind}(P-K)\right]\frac{\Delta(K)}{\omega_n^2+\xi_{\bf k}^2+|\Delta(K)|^2}.
\end{eqnarray}
Here $u_{\rm ff}({\bf p}-{\bf k})$ denotes the instantaneous interaction and $U_{\rm ind}(P-K)$ is the boson-mediated interaction. For $s$-wave pairing and at $T=0$, the gap equation can be expressed as 
\begin{eqnarray}
\Delta(\nu,p)=\int_0^\infty\frac{d\omega}{\pi}\int_0^\infty\frac{k^2dk}{2\pi^2}\left[{\cal K}_{\rm ff}(p;k)+{\cal K}_{\rm ind}(\nu,p;\omega,k)\right]\frac{\Delta(\omega,k)}{\omega^2+\xi_k^2+\Delta^2(\omega,k)},
\end{eqnarray}
with the $s$-wave kernel functions  ${\cal K}_{\rm ind}(\nu,p;\omega,k)$ from the boson-mediated interaction and ${\cal K}_{\rm ff}(p;k)$ from the instantaneous interaction.
It is easy to show that
\begin{eqnarray}
\frac{\partial\Delta(\omega,k)}{\partial\omega}=\int_0^\infty\frac{d\nu}{\pi}\int_0^\infty\frac{p^2dp}{2\pi^2}\frac{\partial {\cal K}_{\rm ind}(\omega,k;\nu,p)}{\partial\omega}
\frac{\Delta(\nu,p)}{\nu^2+\xi_p^2+\Delta^2(\nu,p)}.
\end{eqnarray}
Note that the right-hand side does not depend explicitly on the kernel function ${\cal K}_{\rm ff}(p;k)$. Thus if the magnitude of the gap function is enhanced by the additional instantaneous pairing interaction, the derivative $\partial\Delta/\partial\omega$ is also enhanced.

\section{Results for finite Zeeman field}

For finite Zeeman field $h$, the mean-field grand potential is still given by
 \begin{eqnarray}
\Omega=-\sum_{K,K^\prime}\Delta^*(K)U^{-1}(K-K^\prime)\Delta(K^\prime)-\frac{1}{\beta {\cal V}}\sum_K\ln\det \left[G_h^{-1}(K)\right],
\end{eqnarray}
with the fermion Green's function replaced with
\begin{eqnarray}
G_h^{-1}(K)=
\left[\begin{array}{cc}i\omega_n+h-\xi_{\bf k} & \Delta(K)\\  \Delta^*(K) & i\omega_n+h+\xi_{\bf k}\\ \end{array}\right].
\end{eqnarray}
Minimizing the grand potential, we obtain the gap equation 
\begin{eqnarray}
\Delta(P)=-\frac{1}{\beta {\cal V}}\sum_{K}U(P-K)\frac{\Delta(K)}{(\omega_n-ih)^2+\xi_{\bf k}^2+|\Delta(K)|^2}.
\end{eqnarray}
The density for each spin component is given by
\begin{eqnarray}
n_\uparrow=\frac{1}{\cal V}\sum_{\bf k}n_{{\bf k}\uparrow},\ \ \ \ \ \ \ \ \ n_\downarrow=\frac{1}{\cal V}\sum_{\bf k}n_{{\bf k}\downarrow},
\end{eqnarray}
with the momentum distributions
\begin{eqnarray}
n_{{\bf k}\uparrow}&=&\frac{1}{\beta}\sum_{n}\frac{(i\omega_n+h+\xi_{\bf k})e^{i\omega_n0^+}}{(i\omega_n+h)^2-\xi_{\bf k}^2-|\Delta(K)|^2},\nonumber\\
n_{{\bf k}\downarrow}&=&-\frac{1}{\beta}\sum_{n}\frac{(i\omega_n+h-\xi_{\bf k})e^{i\omega_n0^-}}{(i\omega_n+h)^2-\xi_{\bf k}^2-|\Delta(K)|^2}
=\frac{1}{\beta}\sum_{n}\frac{(i\omega_n-h+\xi_{\bf k})e^{i\omega_n0^+}}{(i\omega_n-h)^2-\xi_{\bf k}^2-|\Delta(K)|^2}.
\end{eqnarray}
For a solution of the gap equation, the grand potential can be expressed as
\begin{eqnarray}
\Omega_{\rm S}-\Omega_{\rm N}=\frac{1}{\beta {\cal V}}\sum_{K}\frac{|\Delta(K)|^2}{(\omega_n-ih)^2+\xi_{\bf k}^2+|\Delta(K)|^2}
-\frac{1}{\beta {\cal V}}\sum_K\ln\left[1+\frac{|\Delta(K)|^2}{(\omega_n-ih)^2+\xi_{\bf k}^2}\right],
\end{eqnarray}

At $T=0$, the gap equation for $s$-wave pairing becomes
\begin{eqnarray}
\Delta(\nu,p)=\int_{-\infty}^\infty\frac{d\omega}{2\pi}\int_0^\infty\frac{k^2dk}{2\pi^2}U_s(\nu-\omega;p,k)\frac{\Delta(\omega,k)}{(\omega-ih)^2+\xi_k^2+\Delta^2(\omega,k)}.
\end{eqnarray}
We can show that $\Delta(\omega,k)=\Delta(-\omega,k)$ still holds for $h\neq0$. Thus the gap equation can be written in a symmetrical form
\begin{eqnarray}
\Delta(\nu,p)=\int_0^\infty\frac{d\omega}{2\pi}\int_0^\infty\frac{k^2dk}{2\pi^2}{\cal K}_s(\nu,p;\omega,k)\left[\frac{\Delta(\omega,k)}{(\omega-ih)^2+\xi_k^2+\Delta^2(\omega,k)}+(h\rightarrow-h)\right].
\end{eqnarray}
Using the same trick of contour integral, the momentum distributions can be evaluated as
\begin{eqnarray}
n_{{\bf k}\uparrow}&=&\frac{1}{2}-\int_0^\infty\frac{d\omega}{\pi}\frac{[\omega^2+\xi_k^2+\Delta^2(\omega,k)-h^2](\xi_k+h)-2h\omega^2}{[\omega^2+\xi_k^2+\Delta^2(\omega,k)-h^2]^2+4h^2\omega^2},\nonumber\\
n_{{\bf k}\downarrow}&=&\frac{1}{2}-\int_0^\infty\frac{d\omega}{\pi}\frac{[\omega^2+\xi_k^2+\Delta^2(\omega,k)-h^2](\xi_k-h)+2h\omega^2}{[\omega^2+\xi_k^2+\Delta^2(\omega,k)-h^2]^2+4h^2\omega^2}.
\end{eqnarray}
Thus the total density and the spin imbalance are given by
\begin{eqnarray}
n_{\rm f}&=&\int_0^\infty\frac{k^2dk}{2\pi^2}\left\{1-2\int_0^\infty\frac{d\omega}{\pi}\frac{[\omega^2+\xi_k^2+\Delta^2(\omega,k)-h^2]\xi_k}{[\omega^2+\xi_k^2+\Delta^2(\omega,k)-h^2]^2+4h^2\omega^2}\right\},\nonumber\\
n_\uparrow-n_\downarrow&=&2h\int_0^\infty\frac{k^2dk}{2\pi^2}\int_0^\infty\frac{d\omega}{\pi}\frac{\omega^2-\xi_k^2-\Delta^2(\omega,k)+h^2}{[\omega^2+\xi_k^2+\Delta^2(\omega,k)-h^2]^2+4h^2\omega^2}.
\end{eqnarray}
If the gap is frequency independent, i.e., $\Delta(\omega,k)=\Delta_k$, a direct integration shows that
\begin{eqnarray}
n_\uparrow-n_\downarrow&=&\int_0^\infty\frac{k^2dk}{2\pi^2}\Theta\left(h-\sqrt{\xi_k^2+\Delta_k^2}\right).
\end{eqnarray}
Therefore, below the Pauli limit the spin polarization is zero. However, for a frequency-dependent gap, the spin polarization becomes nonzero once a Zeeman field is turned on.  
At small $h$, the spin imbalance is well given by $n_\uparrow-n_\downarrow\simeq\chi h$.

\end{widetext}


\begin{thebibliography}{10}
\bibitem{Yukawa-note}   {In this toy model, the scalar mode is not necessary to be the excitation of a relativistic quantum field as in particle physics, 
                                        and thus the velocity parameter $v_\phi$ is not necessary to be equal to the speed of light.}
\bibitem{Note-Dij}         {The details of the derivation of the coefficients $D_{\rm ij}$ will be published elsewhere.}
\bibitem{ABF-note}       {In the recent experiment on $^6$Li-$^{133}$Cs mixture \cite{BF-Mix06}, the inter-species Feshbach resonance is used to tune the boson-fermion scattering length $a_{\rm bf}$. 
                                      The parameters for the stable mixture can be estimated as $n_{\rm f}/n_{\rm b}\sim0.1$, $(n_{\rm b}a_{\rm bb}^3)^{1/3}\sim0.05$, 
                                      and $(n_{\rm b}a_{\rm bf}^3)^{1/3}\sim-0.1$ at the trap center.}
\end{thebibliography}
\end{document}